\documentclass[a4paper]{jpconf}
\usepackage{graphicx}
\begin{document}
\title{Computational neuroanatomy: mapping cell-type densities in the mouse brain, simulations from the Allen Brain Atlas}

\author{Pascal Grange}

\address{Xi'an Jiaotong-Liverpool University, Department of Mathematical Sciences, 111 Ren'ai Rd, Science Building SD557, Suzhou 215123, Jiangsu Province, China}

\ead{Pascal.Grange@xjtlu.edu.cn}

\begin{abstract}
 The Allen Brain Atlas (ABA)  of the adult mouse consists of digitized expression profiles of thousands of genes in the mouse brain, co-registered to a common three-dimensional template (the Allen Reference Atlas). This brain-wide, genome-wide data set has triggered a renaissance in neuroanatomy. Its voxelized version (with cubic voxels of side 200 microns)  can be analyzed on a desktop computer using MATLAB. On the other hand, brain cells exhibit a great phenotypic diversity (in terms of size, shape and electrophysiological activity), which has inspired the names of some well-studied cell types, such as granule cells and medium spiny neurons. However, no exhaustive taxonomy of brain cells is available. A genetic classification of brain cells is under way, and some cell types have been characterized by their transcriptome profiles. However, given a cell type characterized by its transcriptome, it is not clear where else in the brain similar cells can be found. The ABA can been used to solve this region-specificity problem in a data-driven way: rewriting the brain-wide expression profiles of all genes in the atlas as a sum of cell-type-specific transcriptome profiles is equivalent to solving a quadratic optimization problem at each voxel in the brain. However, the estimated brain-wide densities of 64 cell types published recently were based on one series of co-registered coronal {\emph{in situ}} hybridization (ISH) images per gene, whereas the online ABA contains several image series per gene, including sagittal ones. In the presented work, we simulate the variability of cell-type densities in a Monte Carlo way by repeatedly drawing a random image series for each gene and solving optimization problems. This yields error bars on the region-specificity of cell types.\\
{\emph{Prepared for the  International Conference on Mathematical Modeling in Physical Sciences, 5th-8th June 2015,
 Mykonos Island, Greece.}}

\end{abstract}

\section{Introduction} 

The Allen Brain Atlas (ABA, \cite{AllenGenome,AllenAtlasMol}) put neuroanatomy on a genetic basis by releasing voxelized, 
   {\emph{ in situ}} hybridization  data for the expression of the entire genome in the mouse brain  ({\ttfamily{www.mouse-brain.org}}). These data were co-registered to 
the Allen Reference Atlas of the mouse brain (ARA, \cite{ARA}).  About 4,000 genes of special neurobiological interest 
  were proritized. For these genes 
 an entire brain was sliced coronally and processed (giving rise to the coronal ABA). For the rest of the genome 
 the brain was sliced sagitally, and only the left hemisphere was processed (giving rise to the sagittal ABA).\\
 From a computational viewpoint, gene-expression data from the the ABA can be studied collectively,  thousands of genes at a time.
 Indeed the collective behaviour of gene-expression data is crucial for the analysis of \cite{cellTypeBased},
 in which the brain-wide correlation between the ABA and cell-type-specific microarray data was studied.
 These microarray data characterize the transcriptome of $64$ different cell types, microdissected 
 from the mouse brain, and collated in \cite{OkatyComparison}. However, for a given cell characterized in this way,
 it is not known where other cells of the same type are located in the brain. 
A linear model was proposed in \cite{cellTypeBased,supplementary1,supplementary2} (see also \cite{KoCellTypes,TanFrenchPavlidis,JiCellTypes}), and used to estimate
 the region-specificity of cell types  by linear regression with 
  positivity constraint. The model was fitted using the coronal ABA only, which allowed to 
 obtain brain-wide results. However, this restriction implies that only one ISH expression profile per gene
 was used to fit the model. This poses the problem 
 of the error bars on the results of the model.\\

\section{Spatial densities of cell types in the mouse mouse brain from the ABA and transcriptome profiles}

 Since all the ISH data in the ABA were co-registered to the 
 voxelized ARA, so that data for the sagittal and coronal atlas
 can be treated computationally in the same way. However, the ABA does not specify from which cell type(s) the expression of each gene comes.\\
{\bf{Gene expression energies from the Allen Brain Atlas.}} In the ABA, the adult mouse brain is partitioned into $V=49,742$ cubic voxels of side 200 microns, to which ISH data are registered \cite{AllenGenome,AllenAtlasMol,ARA} for thousands of genes.
For computational purposes, these gene-expression data can be arranged into 
 a voxel-by-gene matrix. For a cubic labeled $v$, the {\it{expression energy}} \cite{AllenGenome,AllenAtlasMol}  of the gene $g$ is a
weighted sum of the greyscale-value intensities evaluated at the
pixels intersecting the voxel:
\begin{equation}
E(v,g) = {\mathrm{expression\;energy\;of\;gene\;labeled\;}}g\;{\mathrm{in\;voxel\;labeled\;}}v,
\label{ExpressionEnergy}
\end{equation}
  The  analysis of \cite{cellTypeBased} is restricted to digitized image series from the coronal ABA,
  for which the entire mouse brain was processed in 
 the ABA pipeline (whereas only the left hemisphere was processed for the sagittal atlas).\\
{\bf{Cell-type-specific transcriptomes and density profiles.}} On the other hand, the
 cell-type-specific microarray reads collated in \cite{OkatyComparison} (for $T=64$
 different cell-type-specific samples studied in \cite{foreBrainTaxonomy, ChungCells, ArlottaCells, RossnerCells, HeimanCells,CahoyCells,DoyleCells,OkatyCells}) can be arranged in a type-by-gene matrix denoted by $C$, such that
\begin{equation}
C(t,g) = {\mathrm{expression\;of\;gene\;labeled\;}}g\;{\mathrm{in\;cell\;type\;labeled\;}}t,
\label{typeByGene}
\end{equation} 
 and the columns are arranged in the same order as in the matrix $E$ of expression energies defined in Eq. \ref{ExpressionEnergy}.\\
We proposed the following linear model  in \cite{cellTypeBased} for a voxel-based gene-expression atlas in terms
of the transcriptome profiles of individual cell types and their spatial densities:
\begin{equation}
E(v,g) = \sum_t  \rho_t(v)C( t,g)  + {\mathrm{Residual}}(v,g),
\label{modelEquation}
\end{equation}
where the index $t$ denotes labels cell type, and  $\rho_t(v)$ denotes its (unknown) density at voxel labeled $v$. 
 The values
 of the cell-type-specific density profiles were computed in \cite{cellTypeBased} by minimizing the value
 of the residual term over all the (positive) density profiles, which amounts to solving a quadratic
 optimization problem (with positivity constraint) at each voxel. These computations can be reproduced on a desktop 
 computer using the MATLAB toolbox {\ttfamily{Brain Gene Expression Analysis}} (BGEA) \cite{qbBGEA,BGEAManual}. For other applications
of the toolbox  see \cite{markerGenes} (marker genes of brain regions), \cite{autismCoExpr, autismCoExpr2} for co-expression
 properties of some autism-related genes, and \cite{eikonal} for computations of stereotactic coordinates).\\


\section{Monte Carlo simulation of variability of spatial densities of cell types}

   The optimization procedure in our model is deterministic. On the other hand, decomposing the density of a cell type
 into the sum of its mean and Gaussian noise is 
 a difficult statistics problem (see \cite{Meinshausen2013}).
 Some error estimates on the value of $\rho_t(v)$ were obained in \cite{cellTypeBased}
  using sub-sampling techniques (i.e. 
 sub-sampling the data repeatedly by keeping only a random 10\% of the 
 coronal ABA). This induced
 a ranking of the cell types based on the stability of the results against sub-sampling.
 However, the 10 \% fraction is arbitrary (even though it is close to the 
 fraction of the genome covered by our coronal data set).\\
\begin{figure}
      \includegraphics[width=0.99\textwidth]{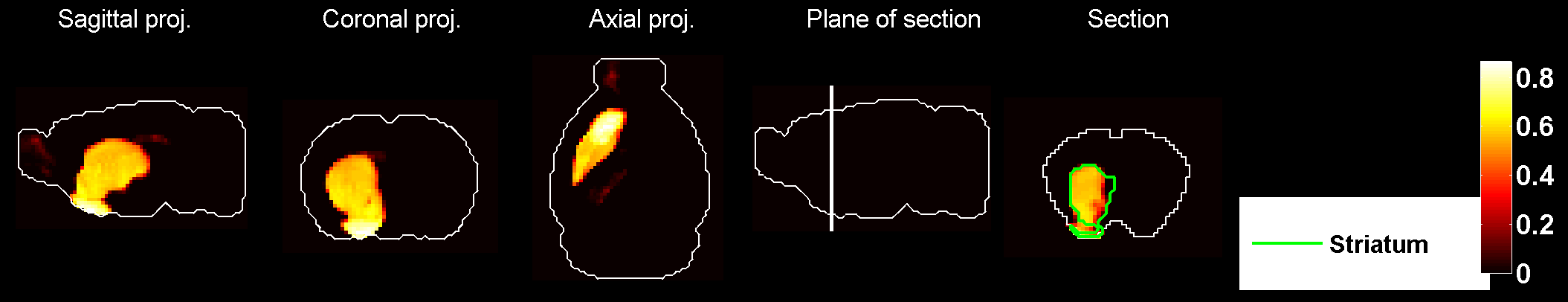}
    \caption{{\bf{Heat map of the average density of cell types in the left hemisphere}},
 $\langle   \rho_t(v) \rangle$, defined in Eq. \ref{meanFittingDef}, for medium spiny neurons, labeled $t=16$ in our data set. The restriction 
 to the left hemisphere comes from the use we made of sagittal image series, which cover the left hemisphere only.}
    \label{meanFittings}
  \end{figure}
In the present work we  simulated the variability of the spatial density of cell types by  integrating the digitized
 sagittal image series into the data set.
  For gene labeled $g$, the ABA  provides $N(g)$ expression profiles, where 
 $N(g)$ is the number of image series in the ABA for this gene. 
 Hence, instead of just one voxel-by-gene matrix,
 the ABA gives rise to a family of $\prod_{g=1}^G N(g)$ voxel-by-gene matrices, with voxels belonging to the 
  left hemisphere.  A quantity computed from the coronal ABA can be recomputed from any of these matrices, thereby inducing
 a distribution for this quantity. This is a finite but prohibitively large number of
 computations, so we took a Monte Carlo 
 approach based on $R$ random choices of images series, described  by the following pseudo-code:\\
{\small{
{\ttfamily{
for all $i$ in $[1..R]$}}\\
{\ttfamily{1. for all $g$ in [1..G], choose an image series labeled by the integer $n_i(g)$ in $[1.. N(g)]$}};\\
{\ttfamily{2. construct the matrix $E_{[i]}$ with entries $E_{[i]}(v,g) =E^{(n_i(g))}(v,g)$,\\
{\ttfamily{3. estimate the density of cell type labeled $t$ using this matrix, call the result $\rho_{t,[i]}$}};\\
{\ttfamily{end\\
}}
}}}}
 The larger $R$ is, the more precise the estimates for the distribution of the spatial  density of cell types will be.
 The only price we have to pay for th e integration of the sagittal ABA is the restriction of the 
 results to the left hemisphere in step 2 of the pseudo-code.

\section{Anatomical analysis of results}
The average 
 density across random draws of image series for cell type labeled $t$ reads:
\begin{equation}
\langle   \rho_t(v) \rangle = \frac{1}{R}\sum_{i=1}^R\rho_{t,[i]}(v).
\label{meanFittingDef}
\end{equation}
\begin{figure}
      \includegraphics[width=1.1\textwidth]{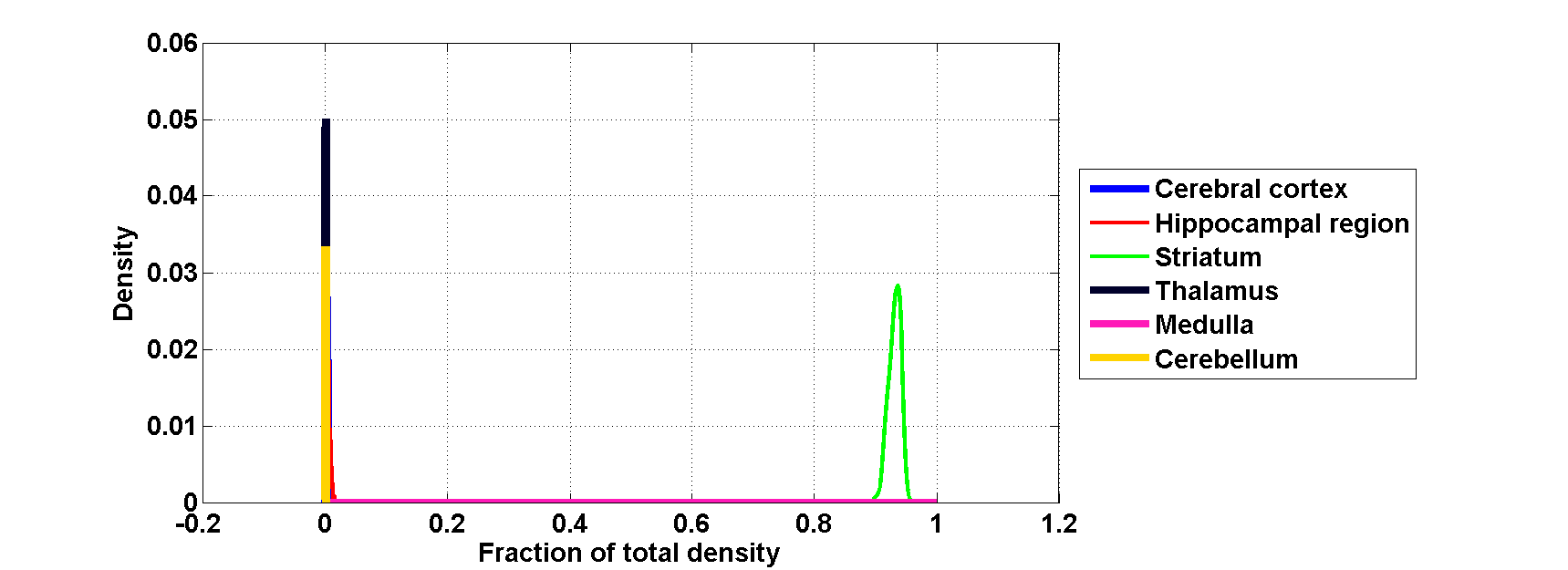}
    \caption{{\bf{Estimated probability densities of fractions of density 
 agglomerated in a few regions of the coarsest versions of the ARA}} (see Eq. 
\ref{fittingDistrDef}), for medium spiny neurons, labeled $t=16$, based on $R=1000$ random draws.
  The right-most peak, corresponding to the striatum, is well-decoupled from the others, furthermore
 the other peaks are all centered close to zero (making most of them almost invisible).  Medium spiny neurons
 have $93(\pm3)$ percent of their densities supported in the striatum,
 without any region gathering more than 5 percent of the signal in any of the random draws.}
    \label{fittingDistrs}
  \end{figure}
A heat map of this average for medium spiny neurons (extracted from the striatum)
 is presented on Fig. \ref{meanFittings}. It is optically very similar to the (left) striatum,
 which allows the model to predict that medium spiny neurons are specific to the striatum (which confirms prior neurobiological knowledge 
 and  therefore serves 
 as a proof of concept for the model).\\
  To compare the results to classical neuroanatomy,  we can group the voxels by region according to the ARA. 
 Since the number of cells of a given type in an extensive quantity, we compute the
 fraction of the total density contributed by a given brain region denoted by $V_r$ (see the legend of Fig. \ref{fittingDistrs} for a list of possible values of 
 $V_r$):
\begin{equation}
 \phi_{r,[i]}(t)  =  \frac{1}{\sum_{v\in\mathrm{left\;hemisphere}}\rho_{[i],t}(v) }\sum_{v\in V_r} \rho_{t,[i]}(v).
\label{fittingDistrDef}
\end{equation}
We can plot the distribution of these $R$ values for a given cell type and all brain regions (see Fig.
 \ref{meanFittings} for medium spiny neurons, which gives rise to the best-decoupled right-most peak
 in the distribution of simulated densities). Moreover, we estimated the densities of the contribution of each region in the coarsest version 
 of the ARA to the total density of each cell type in the data set. For most cell types, this
  confirms the ranking of cell types by stability obtained in \cite{cellTypeBased}, but based on error bars
 obtained from the same set of genes in every fitting of the model (see the accompanying preprint \cite{accompanying}
 for exhaustive results for all cell types in the panel). The most stable results against sub-sampling 
 tend to correspond to cell types for which the anatomical distribution of results is more peaked. The present analysis can be repeated 
 when the panel of cell-type-specific microarray expands.\\

\ack{Microarray data were made available by Ken Sugino, Benjamin Okaty and Sacha B. Nelson. The Allen Atlas data were analysed under the guidance of Michael Hawrylycz and Lydia Ng. This work is supported by the Research Development Fund and the Research Conference Fund of Xi'an Jiaotong--Liverpool University.}


\end{document}